\documentclass[10pt, conference, letterpaper]{IEEEtran}

\usepackage[cmex10]{amsmath}
\usepackage{amssymb}
\usepackage{cite}
\usepackage{graphicx}
\usepackage{epsfig}
\usepackage{url}
\usepackage{color}
\usepackage{setspace,epstopdf,subfigure}
\usepackage{verbatim}
\usepackage{multirow}
\usepackage{array}
\usepackage{float}

\newcommand{\ignore}[1]{}

\usepackage[belowskip=-1pt,aboveskip=5pt]{caption}

\newcommand\MyBox[2]{
  \fbox{\lower0.75cm
    \vbox to 1.7cm{\vfil
      \hbox to 1.7cm{\hfil\parbox{1.4cm}{#1\\#2}\hfil}
      \vfil}%
  }%
}

\begin{document}

\title{FlowIntent: Detecting Privacy Leakage from User Intention to Network Traffic Mapping}

\author{
    \IEEEauthorblockN{Hao Fu, Zizhan Zheng, Aveek K. Das, Parth H. Pathak, Pengfei Hu, Prasant Mohapatra}
    \IEEEauthorblockA{Department of Computer Science, University of California, Davis, CA, USA.
    \\\texttt{\{haofu, cszheng, akdas, phpathak, pfhu, pmohapatra\}@ucdavis.edu}}
}

\maketitle

\begin{abstract}
The exponential growth of mobile devices has raised concerns about sensitive data leakage.
In this paper, we make the first attempt to identify suspicious location-related HTTP transmission flows from the user's perspective, by answering the question: 
\emph{Is the transmission user-intended}?
In contrast to previous network-level detection schemes that mainly rely on a given set of suspicious hostnames, our approach can better adapt to the fast growth of app market and the constantly evolving leakage patterns. 
On the other hand, compared to existing system-level detection schemes built upon program taint analysis, where {\it all} sensitive transmissions as treated as illegal, our approach better meets the user needs and is easier to deploy. 
In particular, our proof-of-concept implementation (FlowIntent) captures sensitive transmissions missed by TaintDroid, the state-of-the-art dynamic taint analysis system on Android platforms. Evaluation using 1002 location sharing instances collected from more than 20,000 apps 
shows that our approach achieves about 91\% accuracy in detecting illegitimate location transmissions.
\end{abstract}

\section{Introduction}\label{sec:intro}
Smart phones are becoming indispensable to many of us, thanks to the rich functionalities provided by a large number of mobile applications (or apps, for short). For instance, more than 1,500,000 apps can be downloaded from Google Play, the major Android application market in the U.S.\cite{numofapps}. The sheer number of these apps, however, makes it challenging to understand their behavior and control their quality before publishing. Given that many of these apps can access sensitive user data, such as location, contact information and media files, they can potentially share these data in an unintended way, which compromises the user's privacy.
Therefore, it is important to design an automatic framework to detect privacy leakage caused by mobile apps.

\begin{figure}
\centering
\includegraphics[width=0.5\textwidth, height=2.0in]{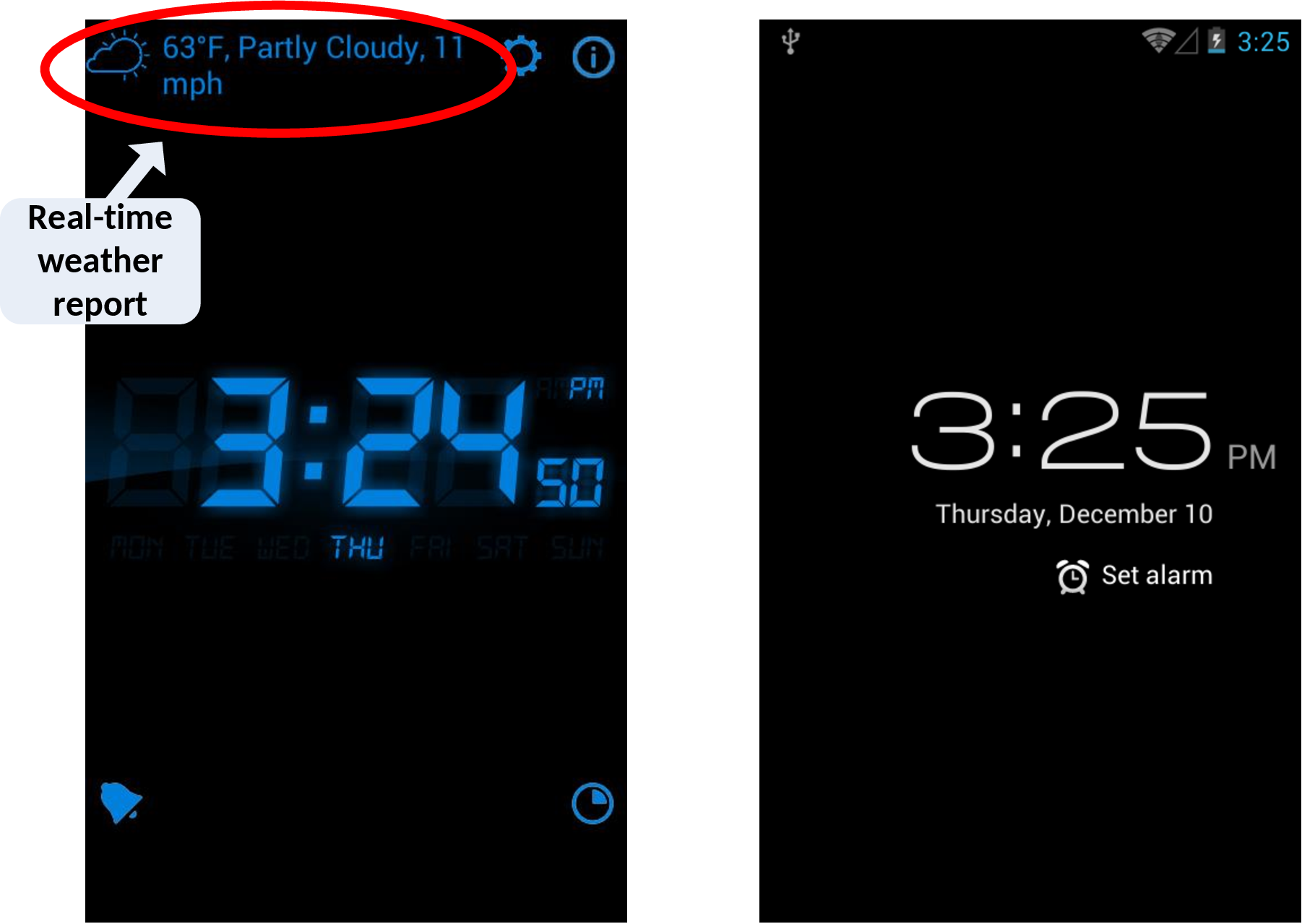}
\caption{Two UI windows of running alarm clock apps}
\label{Fig.UI}
\end{figure}

In this paper, we focus on detecting privacy leaking transmissions from authentic apps in popular Android markets.
Compared to malware, authentic apps still perform their proposed functionality.
However, they may also put users at risk by behaving in a user-unexpected way, such as stealthily sending user's private information out for purposes such as  analytics, advertising, cross-application profiling, and social-computing~\cite{tripp2014bayesian}.
To address this critical problem, we take into account {\it user intention} and define illegal flows as follows:
\textbf{a transmission flow is illegitimate if given the app-level context, the user considers the transmission is not required for its functionalities.}
The app-level context includes both static features that can be obtained without launching the app such as app name and app description, and dynamic features acquired during runtime such as the user interface (UI) that a user is interacting with. Static features provide the general information about the main functionality of an app to help label its transmissions.
For example, a location-sharing flow generated from an alarm clock app is likely illegal since alarm apps typically do not need location information to fulfill their functionality. Dynamic features are also necessary to distinguish between apps of similar functionality, and the different contexts of the same app.
Figure~\ref{Fig.UI} illustrates two UI windows of running alarm clock apps.
The location-sharing flows generated under the first window have a higher chance of being legitimate due to the weather forecast related symbols. 


Our work is the first attempt to identify the above mentioned illegal transmissions with network traffic monitoring. Compared with system-level solutions, our approach is easier to deploy and incurs less overhead at the user side. For instance, we can deploy it at an Intrusion Detection System (IDS) or an access point, and utilize existing network infrastructures to monitor a large number of traffic flows generated by different apps and devices, therefore removes the user-side overhead~\cite{nsdi-clustering}.
Moreover, as we show later in the paper, through a proper modeling of user intention, our approach can potentially capture illegal transmissions missed by system-level approaches.

The state-of-the-art network traffic based approach~\cite{raghuramu2015uncovering} relies on existing lists of malicious hostnames, such as VirusTotal\cite{virustotal}. However, generating such lists often requires significant human effort, which is especially challenging for mobile platforms since a large number of apps are normal apps that only leak private data occasionally. Moreover, manually generated lists of ad, analytic and malicious hostnames can barely keep pace with the fast growth of app market, where new hostnames are continuously mushrooming. Further,
it is often difficult to tell even for humans whether a flow is targeting an unexpected destination, simply from the hostname.
This is especially true when the illegal flows share the same hostnames with the legal flows. 
Through a proper modeling of user intention, our approach largely relaxes the dependence on human intervention, which can potentially lead to fast generation of signatures for new traffic patterns.

In this paper, we present FlowIntent, a proof-of-concept system that generates signatures for identifying user-unintended leakage in HTTP flows. FlowIntent focuses on location related leakage currently, but our approach can potentially be extended to detect other types of privacy leakage as well. Our general approach works as follows. From the set of apps crawled from app markets, we first identify the subset of apps that share locations. Those apps are then automatically launched and executed for a short period of time, with corresponding user interface and data traffic recorded.
Each $\langle app, window \rangle$ pair is then treated as an app-level context of a running instance, and is labeled as either ``expected'' or ``unexpected'' in terms of location transmission, based on the visible app meta-data including app name, description and the UI for the running period.
These data are used to train a user intention model, which helps to classify new running instances automatically. All the location-sharing HTTP flows generated by ``unexpected'' instances are then labeled as illegal flows, while all the location sharing flows from ``expected'' instances are labeled as legal flows.

We note that some flows from ``expected'' instances may actually target a malicious receiver, and should be classified as illegal transmissions, even if the type of sensitive data they carry does not violate the context. We leverage existing hostname lists of ad and analytics servers to identify (a part of) illegal flows
from the ``expected'' instances, so that our approach can be easily made automatic. We get the ground truth for all the flows manually in the testing stage. 
Our testing results show that even if the training data may contain some mislabelled flows, supervised learning can still achieve accurate results through user intention modeling and existing hostname lists, which indicates that the mislabelled traffic is limited and does not have a significant impact on the classification result.
We have further implemented an unsupervised learning approach that only relies on the illegitimate traffic generated by ``unexpected'' instances, which is less accurate than the supervised approach as expected, but incurs less labeling overhead.
FlowIntent implements both approaches, which enables the administrators to switch between them to meet their specific requirements. The network signatures generated by FlowIntent can then be deployed at IDS to detect user-unintended location-sharing HTTP flows and warn the corresponding users.

We emphasize that the objective of FlowIntent is to identify privacy leaking flows instead of leaking apps. The app-level information is only used to train the user intention model, and the network-level signatures generated by FlowIntent are purely based on the characteristics of HTTP flows. Therefore, IDS does not need to know which app has generated a flow in order to identify privacy leakage in the flow. Our work is orthogonal to existing works that focus on learning app identity from network traffic such as NetworkProfiler\cite{dai2013networkprofiler}.

We have evaluated FlowIntent on location sharing flows from 1002 instances extracted from more than 20,000 apps crawled from Android app markets. Our main contributions can be summarized as follows.
\begin{itemize}
\item We propose a more accurate definition of privacy leakage in mobile platforms that combines both app level behavior (e.g, does the user intend to share the location information) and flow level behavior (e.g., where the location information is sent to).
\item We develop a lightweight approach for detecting privacy leakage in mobile platforms from network traffic data. Our approach is easy to deploy and can adapt to the constantly evolving leakage patterns.
\item Our approach achieves about 91\% accuracy on the 1002 location leaking instances. Moreover, our learning model is able to identify location sharing flows that are missed by \textit{TaintDroid}\cite{enck2014taintdroid}, the state-of-the-art dynamic taint analysis tool for Android platforms, which indicates the benefit of considering traffic level features.
\end{itemize}

The rest of the paper is organized as follows. We present related work in Section~\ref{sec:related}, followed by an overview of the system in Section~\ref{sec:system} . We discuss our user intention modeling and traffic learning schemes in Sections~\ref{sec:ui} and~\ref{sec:learning}, respectively. After presenting the evaluation results in Section~\ref{sec:eval}, 
we conclude our paper in Section~\ref{sec:conclusion}.

\begin{figure*}[ht]
\centering
\includegraphics[width=0.65\textwidth, height=2.0in]{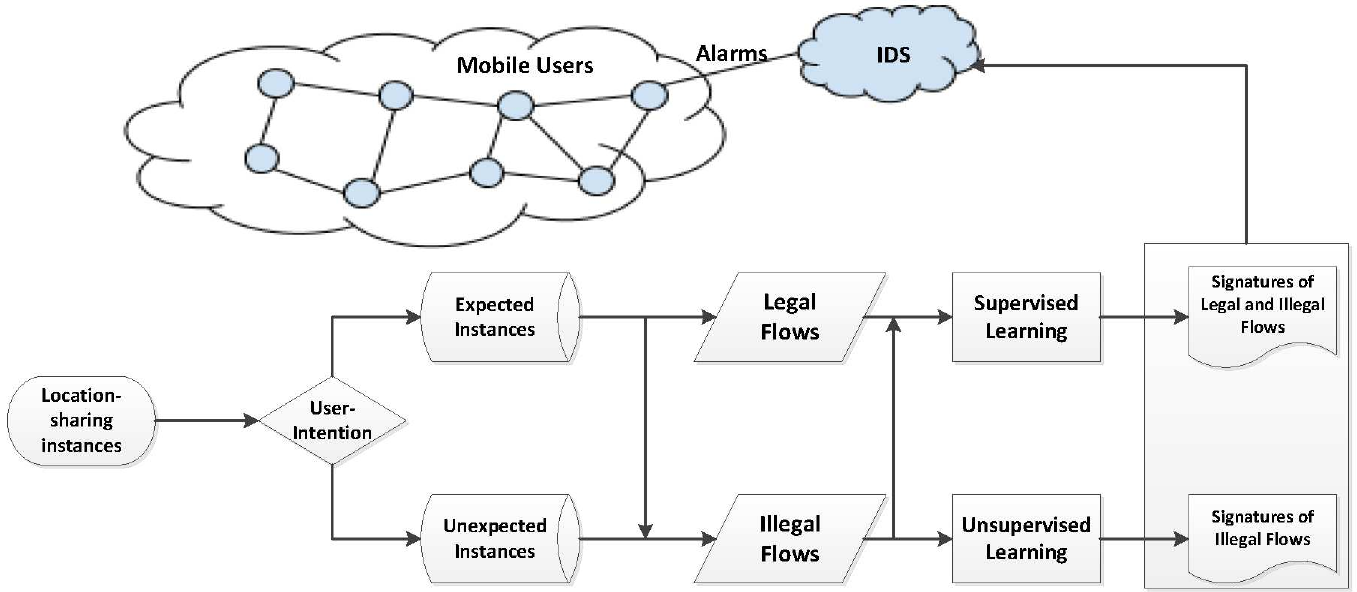}
\caption{FlowIntent System Architecture}
\label{Fig.arch}
\end{figure*}

\section{Related Work}\label{sec:related}

\noindent{\bf Program taint analysis}: Static and dynamic program taint analysis of apps focus on identifying whether sensitive data leaves the user device. TaintDroid~\cite{enck2014taintdroid} 
tries to resolve privacy leaking by modeling the behavior of app through a dynamic analysis system, whereas FlowDroid~\cite{arzt2014flowdroid} and IccTA ~\cite{li2015iccta} adopt static approach to model the app behavior through byte code inspection. They treat all sensitive transmission as illegal, so that suffer from high false positive rate.

\noindent{\bf User intention modeling}: AppIntent~\cite{yang2013appintent} first proposes the problem of the mismatch between user intention and app behavior by integrating symbolic execution, static taint analysis and dynamic analysis. However, this work is not fully automatic since they require users to manually specify their expectations. Whyper~\cite{pandita2013whyper} and AotuCog~\cite{qu2014autocog}  model user expectation through app's description and check it with permissions and API. Gorla et al. follows the same idea but replaces permissions to API calls~\cite{gorla2014checking}. We argue that both permissions and API calls are coarse grained and do not reflect real run-time behaviors.
AsDroid~\cite{huang2014asdroid} examines mismatch between keywords on limited visible buttons and underlying codes.
Their static analysis lacks enough contextual information to present more accurate results and the high analysis overhead makes it impossible to be deployed at devices. Such drawbacks render it impractical for use from the user's perspective.

\noindent{\bf Network traffic based approaches}: Another research thread that is relevant to our work is internet traffic monitoring, which is used for network traffic classification and determining different protocols and applications being used by the users. As discussed in~\cite{demystify} (and references therein), traffic monitoring and classification methods can be used for anomaly detection~\cite{anomaly1}, location categorization~\cite{localization} and also for malware detection. A number of research works have focused on detection of malicious traffic from network data (or specifically HTTP traces and URLs). These works have used a number of different approaches, like clustering~\cite{nsdi-clustering} or keyword-based lexical features~\cite{url-lexical, raghuramu2015uncovering}. However, in our work, we are focused on automatically identifying privacy disclosure caused by the authentic apps with the help of user intention modeling, instead of identifying malicious traffic generated from certain malware families.


\section{System Overview}\label{sec:system}
In this section, we present a high-level overview of our approach for detecting user unintended sensitive data transmissions. We focus on location transmissions since it is one of the most common types of privacy leakage on mobile platforms\cite{xia2015effective}. However, we expect that our general approach can be extended to other types of privacy leakage.

Figure~\ref{Fig.arch} gives an overview of our approach, which works as follows:

\begin{itemize}
\item {\bf Data Collection:} From the set of apps crawled from Android app markets, we identify the subset of apps that share locations with the help of dynamic taint analysis, and collect running instances from them.
For each running instance, we store its app-level contextual data, along with the captured location-sharing traffic flows.
\item {\bf User Intention Modeling:} We build an user intention model for location sharing instances using supervised learning (Section~\ref{sec:ui}). A location sharing instance is labeled ``unexpected'' if it is not supposed to send location data according to its app-level context, and is labeled ``expected'' otherwise.
The purpose of user intention modeling is to help in identifying sets of legitimate and illegitimate flows to build the traffic models in the next step. 
\item {\bf Traffic Classification:} We then apply machine learning to build traffic models for location sharing transmissions, using both statistical and lexical features in traffic data (Section~\ref{sec:learning}). 
Two learning approaches are considered. First, we use supervised learning to classify 
each flow into legitimate location transmissions, illegal location transmissions, and non-location transmissions.
Supervised learning provides high accuracy and a good characterization of location sharing behavior of mobile apps, but also incurs high overhead as it requires more 
human intervention to label the flows in the training stage. 
Second, we use unsupervised learning to build a model purely based on illegitimate flows, and treat both benign location transmission and non-location transmissions as outliers. Although its prediction results are less accurate compared with the supervised approach, unsupervised learning only requires a set of illegitimate flows, 
reducing human intervention. 
\end{itemize}

It is important to note that for traffic modeling, only network traffic data is used in the testing stage, while user intention modeling and app-level information are only applied in the training stage. 
Therefore, our approach can be implemented at IDS to remove user-side overhead, while still achieving good performance by taking advantage of user intentions modeling.

\begin{figure*}[ht]
\centering
\includegraphics[width=0.8\textwidth, height=1.6in]{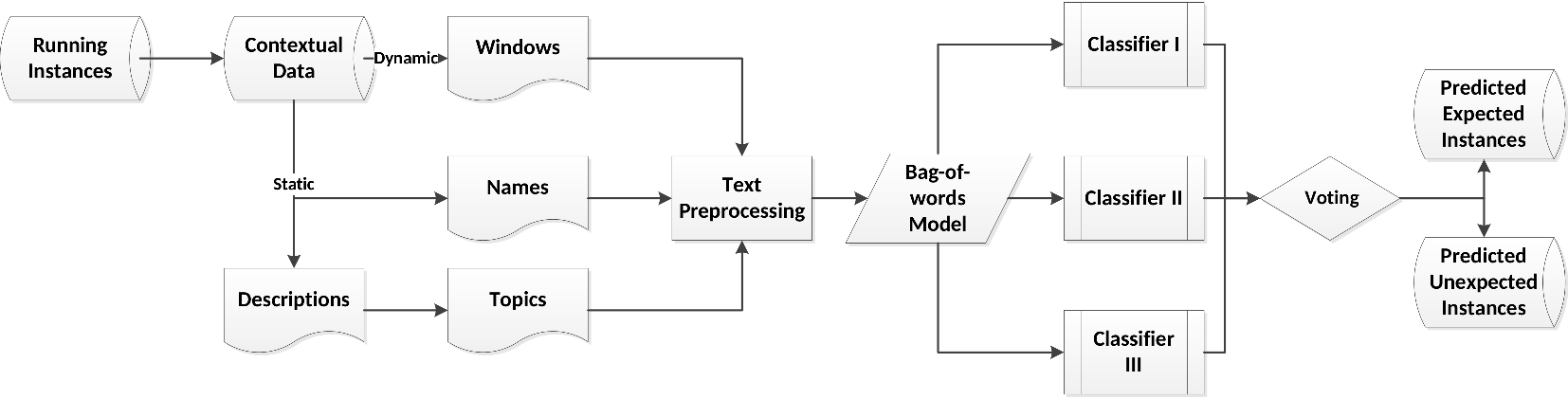}
\caption{User Intention Modeling}
\label{Fig.uiarch}
\end{figure*}

\section{User Intention Modeling}\label{sec:ui}
As we discussed before, it is important to understand user intention to distinguish between benign and illegitimate location sharing behavior. A proper modeling of user intention, however, is challenging. Intuitively, a user's intention for a specific transmission of sensitive data can be faithfully captured by his/her knowledge about the app's general functionality, and the sequence of windows and input/output right before the transmission. 
It is therefore reasonable to model user intention regarding a sensitive transmission from the app-level context when the transmission is generated, including both static features that are fixed to the app and can be acquired without launching the app 
and dynamic features that vary over time/input and can be only obtained during runtime. 
As a first step towards this general model, we narrow the problem scope and consider a simplified approach in this paper. We define a running instance as a snapshot of a running app that involves a single window, and ignore user interactions. Each running instance is then associated with an app-level context that is represented by an $\langle app, window \rangle$ pair, where the \textit{app} component denotes the static features of the app including app name and description, and the \textit{window} component denotes the corresponding window of the specific running instance. We note that an \textit{app} component may be associated with several different \textit{window} components. Extension of our modeling to consider multiple correlated windows and user interactions is part of our future work.

We focus on text related features, and utilize text classification, which is commonly used in Natural Language Processing (NLP) and spam-detection~\cite{forman2003extensive}, to distinguish between the legal and illegal location transmissions. The overall architecture of the user-intention modeling module is shown in Figure~\ref{Fig.uiarch}.
The output of this step is a classification model that can distinguish between ``expected'' and ``unexpected'' sharing instances. 


\subsection{Text Preprocessing and Feature Extraction}
In this subsection, we discuss in detail each type of features derived from text-based metadata of apps that we use for modeling user intentions.

\noindent{\bf Static Contextual Features}: Static features provide general information about the {\it main} functionality of an app that can help label its transmissions. In this work, we consider both app descriptions and app names as static features.

For app description, we follow the approach in~\cite{gorla2014checking} to map app descriptions into topics, which provides a concise representation of the apps' main functionalities. The approach works as follows.
\begin{itemize}
\item We first utilize the \textit{Natural Language Toolkit} \cite{nltk} in python to tokenize English sentences in a description into words. The words are then fed to the stemmer, where they are reduced to their root forms. We also remove all the stop words. We do the similar things for apps with Chinese descriptions by using \textit{Jieba}\cite{jieba}.
\item We then apply text mining to get the most related topics. Since detailed topic modeling is beyond the scope of this paper, we directly leverage the set of keywords for each corresponding topic given in~\cite{gorla2014checking}. For a description that includes keywords belonging to different topics, we choose the topic that is hit by the maximum number of different keywords. 
\item It is possible that a description does not fall into any topics generated by~\cite{gorla2014checking}. We assign such an app a coarse topic given by the corresponding app market.
\end{itemize}

Each topic is treated as a single feature in our learning model. Table~\ref{tab:topics} gives some examples of topics given in~\cite{gorla2014checking}, Google Play, and Baidu App Market. Table II shows the keywords associated with some topics.

We also consider app names as our static contextual features. The intuition is that it is likely that developers will give their apps names to match their functionalities. This is especially true for apps in the ``tools'' category, which are usually named based on their core functionalities. For instance, we expect that the app ``LocalWeather'' to be a weather app and ``SuperLed'' to be a flashlight app. We do not directly use an app name as a feature. Instead, we extract popular words that frequently appear in app names with the help of existing word list~\cite{google-english}. 
For instance, for ``LocalWeather'', ``Local'' and ``Weather'' are extracted. We also add some new words such as ``tech'' and ``nav''.
We have generated 260 binary features from app names in total.

\begin{table}
\caption{Sample Topics}
\label{tab:topics}
\resizebox{\columnwidth}{!}{%
\begin{tabular}{ l | r }
\hline	
& personalize, games and cheat sheets, music,\\
Topics from \cite{gorla2014checking} &  navigation and travel, language, share, health,\\
& kids, ringtones and sound, search and browse\\
\hline
& sports, social, shopping, productivity,tools,\\
Google Play & photography, personalization, medical, lifestyle,\\
& finance, libraries and demo, music and audio\\
\hline
& social and communication, system and tools,\\
Baidu App Market & finance and shopping, themes, photography,\\
& video and audio, lifestyle, office, books \\
\hline
\end{tabular}
}
\end{table}

\begin{table}
\label{tab:topic-keyword}
\caption{Example topics with relevant keywords~\cite{gorla2014checking}}
\resizebox{\columnwidth}{!}{%
\begin{tabular}{ l | r }
\hline	
``navigation and travel'' & map, inform, track, gps, naving, travel, citi \\
``weather and stars'' & weather, forecast, locate, temperatur, city, light\\
``health'' & weight, bodi, exercise, diet, workout, medic \\
\hline
\end{tabular}
}
\end{table}

\noindent{\bf Dynamic Contextual Features}: Dynamic contextual features such as the UI provide hints of the runtime behavior of an app. For instance, run-time windows not only help distinguish between similar apps, but also help differentiate running instances of the same app.
Unlike a description that is generally constructed with complete sentences, the text content of a UI component is often presented by phrases or short incomplete sentences. Therefore, it is possible to achieve accurate classification without using sophisticated NLP techniques that is needed to preserve the order of the tokens. In this work, we utilize the simple yet powerful bag-of-words technique that is commonly used in spam detection by treating each distinct word appeared in the window of a running instance as a separate (binary) feature. We extract the text content in the UI components dumped by \textit{UiAutomator}~\cite{uiautomator}, a standard tool supported by Google to extract UI features. However, UiAutomator may miss some text information, especially those related to the web content. 
When this happens, we manually extract the relevant text content from the screen shot of the window. 
In addition to pure text content, we have added a binary feature ``city-clickable'' to represent a special type of clickable widgets. Some user-expected instances have a single checkable button named with a city name shown in the window. Those instances take user's location to deliver region related news or services. Users can change their regions by clicking that button. On the other hand, we seldom observe such implementation on ``unexpected'' cases.

\begin{figure}
\centering
\includegraphics[height=1.6in,width=0.5\textwidth]{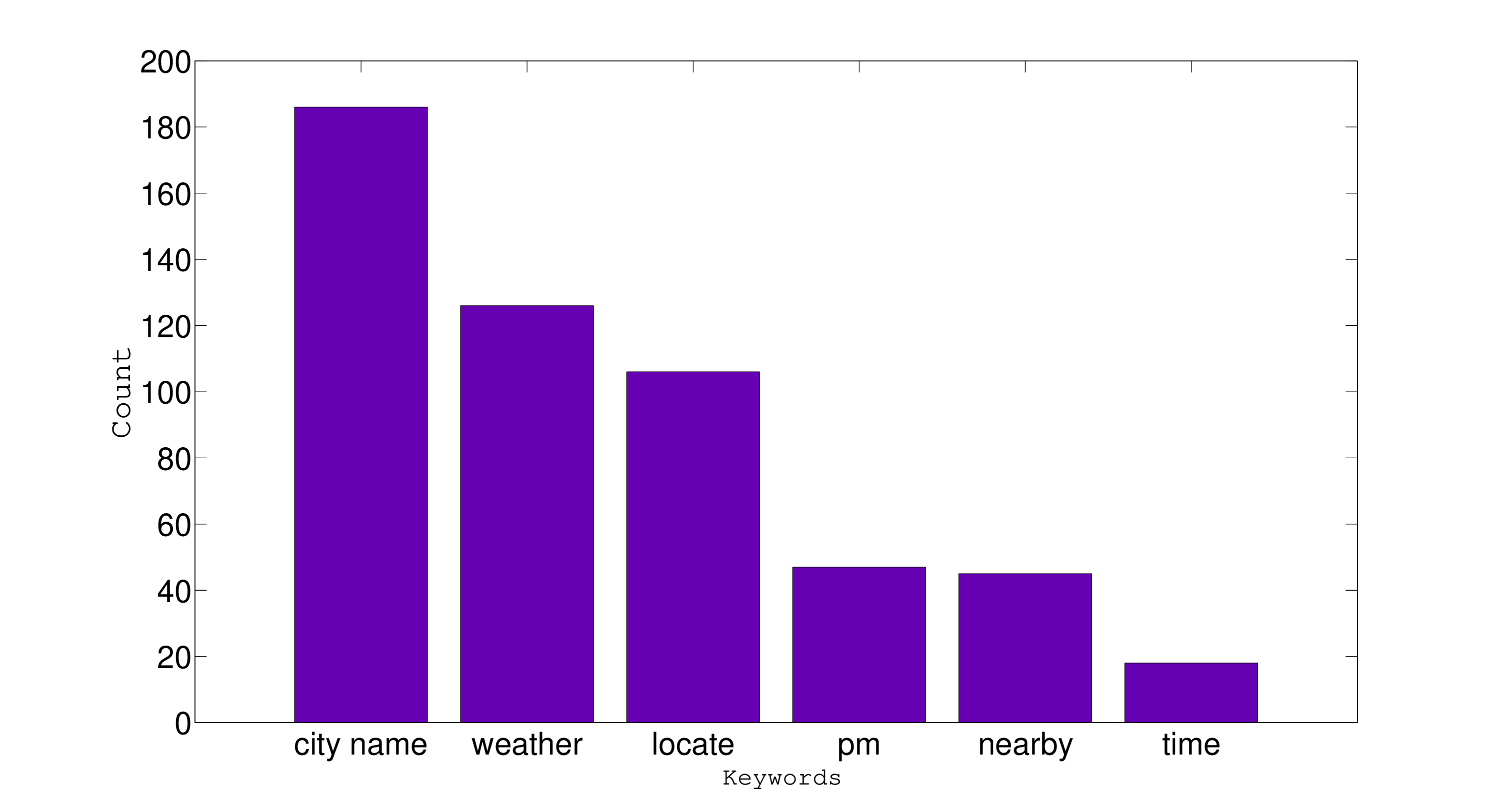}
\caption{Popular keywords of UI in 634 legal local-sharing instances}
\label{Fig.ui_keywords}
\end{figure}

Figure~\ref{Fig.ui_keywords} shows the set of most popular tokens that appear in the UI of legal location-sharing applications, where ``city name'' includes all the concrete city names, and ``pm'' indicates either particulate matter or post meridiem. We observe that these keywords closely match our intuition about location related instances. In particular, ``city name'', 
``locate'' and ``nearby'' are directly related to locations, 
while ``weather'' and ``pm'' appear in weather reporting instances, which are typically location sensitive. Moreover, many location related services have a small widget that shows the current time, which explains why ``time'' is also a popular keyword.  
We have generated 4808 binary features for UI in total.

\subsection{Learning Method}
As a first order approximation of app behavior, we consider one running instance per app that corresponds to the front-page window of the app (see Section~\ref{sec:eval} for details). Although this is a simple approach, it can actually capture a significant amount of suspicious behavior. As observed in~\cite{crussell2014madfraud}, unintended ad and analytics behaviors 
often happen once an app starts. Our approach can be readily extended to consider multiple instances from the same app, where each instance may contain multiple windows. The main challenge, however, is to make this process automatic and representative of the usage patterns of real users, which largely remains open and is part of our future research.

The ultimate purpose of user-intention modeling is to generate (part of) the training data for our traffic classification model to be discussed in the next section. In particular, in the training stage of our traffic modeling, all the location-sharing HTTP flows generated by ``unexpected'' instances are labeled as illegal, while all the location sharing flows from ``expected'' instances are {\it initially} labeled as legal. Therefore, it is important to ensure a high accuracy at this stage. 
To get more precise results, we consider three learning algorithms, random forest, naive Bayes and logistic regression, and adopt a commonly used consensus voting approach~\cite{mislabeled-1999} to filter potential misclassified instances. That is, only the set of instances where all the three algorithms give the same classification results are retained. We provide the evaluation results of our user intention model in Section~\ref{sec:eval-ui}.

\section{Traffic Flow Learning}\label{sec:learning}
In this section, we describe our traffic level features and learning models for detecting illegal traffic flows. 
We consider two learning models: (i) Supervised learning model that generates signatures for legal location flows, illegal location flows, and non-location flows; (ii) Unsupervised approach that builds a model for illegal flows only. As discussed later, the two approaches provide a different accuracy vs. labeling overhead tradeoff.

\subsection{Flow-Level Features}\label{sec:features}
To build traffic learning models that can predict illegitimate location sharing from network-level signatures, we consider both \textbf{statistical features} and \textbf{lexical features}. 

\noindent{\bf Statistical features}
For each HTTP flow that forms a session and is identifiable by a 4-tuple $<$source IP, source port, destination IP, destination port$>$, the following statistical features are calculated:
\begin{itemize}
  \item Total number of TCP packets
  \item Total number of uplink TCP packets
  \item Total number of HTTP packets (Packets with HTTP application layer present)
  \item Packet size of all TCP packets
  \item Packet size of uplink TCP packets
  \item Packet size of downlink TCP packets
  \item Time interval between two consecutive TCP packets
\end{itemize}

\begin{figure*}[t]
\centering
\subfigure[Total number of TCP Packets]{
\label{Fig.tcpcount}
\includegraphics[height=1.8in,width=2.2in]{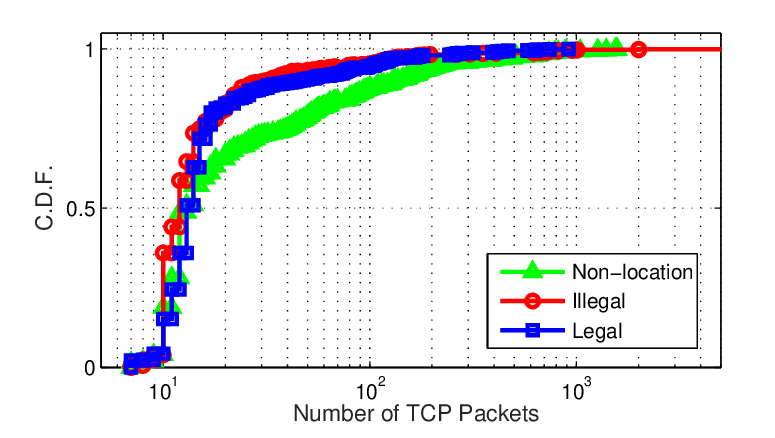}}
\subfigure[Max length of downlink TCP packets (bytes)]{
\label{Fig.maxdown}
\includegraphics[height=1.8in,width=2.2in]{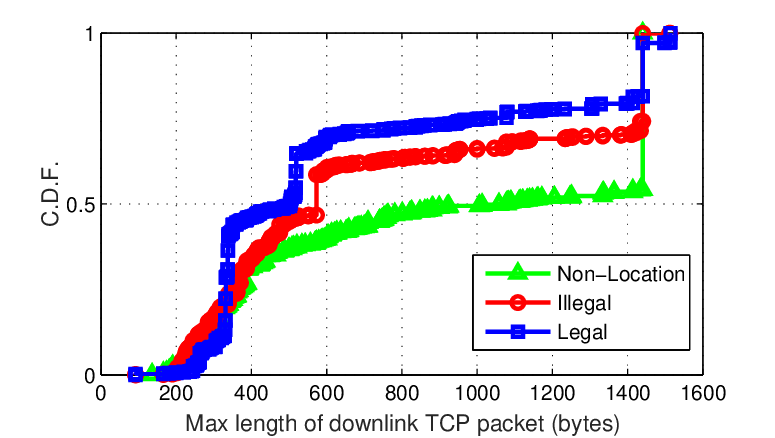}}
\subfigure[ Time interval between two consecutive TCP packets (ms)]{
\label{Fig.interval}
\includegraphics[height=1.8in,width=2.2in]{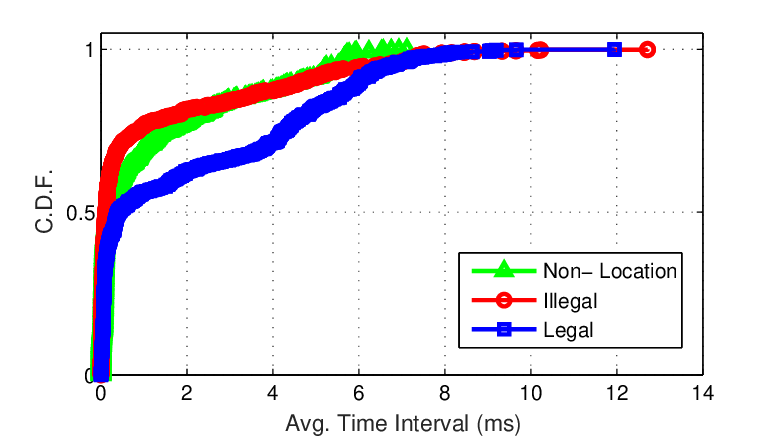}}
\caption{CDFs of Statistical Features }
\label{Fig.stats}
\end{figure*}

The first three attributes hold single numerical values for the count, while the rest four attributes represent a distribution which is represented using 7 statistical features, namely, minimum, maximum, median, mean, standard deviation, skewness and kurtosis. While extracting the features from the traffic, no prior information about the shape of the distributions (Gaussian or not) are known. Since our primary concern is the accurate representation and description of distributions obtained from the data, similar to the network features used in~\cite{demystify}, we consider the first four moments (mean, variance, skewness and kurtosis) in addition to maxima, median and minima. The total number of statistical features is 31.

To understand how these statistical features contribute towards distinguishing between non-location flows, legal location usage and illegal usage flows, three representative characteristics are shown in Fig.~\ref{Fig.stats}. Fig.~\ref{Fig.tcpcount} shows that the non-location flows have a significantly higher number of TCP packets as compared to the location flows, which is expected, while illegal location flows have a slightly lower packet count than legal flows. 
Fig.~\ref{Fig.maxdown} shows that non-location flows have larger downlink packet sizes compared location flows, mainly contributed by data-intensive apps. On the other hand, illegitimate flows are typically ad flows that are usually responded with an advertisement, which makes the maximum packet size in the downlink traffic larger than those for legal flows, where the response is mainly control packets denoting that the location has been received. Fig.~\ref{Fig.interval} shows that the legal flows have higher average packet inter-arrival time,
since the benign servers need more time to handle users' specific location-related requests and generate corresponding responses.

\noindent{\bf Lexical features}
In addition to statistical features, we also consider lexical features derived from the textual properties of URLs, which often contains useful patterns to distinguish benign and malicious traffic~\cite{url-lexical}. The intuition is that location-sharing URLs may contain words that can be used to differentiate the purposes of location requests. As an example, consider an illegitimate location sharing with the following URL: \emph{ads.appsgeyser.com/?\&guid=a5141e1d\&tlat=38.53203\&tlon=-121.759603\&p=android\&test=1}.
We can see that the domain name \emph{ads.appsgeyser.com} has a prefix of ``ads'', which indicates the advertisement purpose of the request.
As another example, consider a location-sharing flow generated by a weather forecast application with the URL \emph{v.juhe.cn/weather/geo?\&lon=-121.750683\&lat=38.540323}. The path portion of the URL \emph{weather/geo?\&lon=-121.750683\&lat=38.540323} includes the word ``weather'', indicating that the server behind the URL is a weather information provider. Moreover, both URLs contain exact longitude and latitude values in the plain text, which can used to identify location-sharing flows from all outgoing traffic traces.
We follow the ``bag-of-words'' approach used in~\cite{url-lexical} and treat each token inside a URL as a binary feature. Below is a list of lexical features considered.
\begin{itemize}
  \item Binary feature for each token in the host name and in the path URL
  \item Length of the host name and entire URL
  \item Number of dots in the URL
\end{itemize}

\subsection{Flow Dataset}\label{sec:dataset-flow}
As we discussed earlier, all location sharing flows from the ``unexpected'' instances are labeled as illegitimate. However, we cannot treat all the flows generated by ``expected'' instances as legitimate. In particular, although for these instances, location sharing is allowed in the current context, the data may be sent to an malicious receiver. To address this problem, we leverage existing hostname lists of ad and analytics servers~\cite{crussell2014madfraud} to identify (a part of) illegal flows from the ``expected'' instances. 
We note that some of these flows are possibly mislabelled due to (1) the inaccuracy of the user intention model, and (2) some of the illegal flows from the ``expected'' running instances are missed as they do not match the existing hostname lists. We allow these noise in the training data, but manually obtain the ground truth of all the flows in the testing stage as discussed below.


To get the ground truth of each flow in the testing stage, the following steps are applied: (1) We first query the ground truth of each running instance 
and label all the flows generated by the true ``unexpected'' instances as illegal; (2) For each flow generated by the true ``expected'' instances, we first exam the destination hostname of the flow. If it belongs to an advertisement or analytics company, the flow is labeled as illegitimate;
(3) We then check the plain text content in the response, 
and the flow is labeled as illegitimate if the response is unrelated to the location sent;
(4) For the rest of flows that cannot be determined by above approaches, 
we have implemented a blocking approach as follows. For each of these flows, we first set firewall rules (based on the TaintDroid reports) to block the flow.
We then clean the cache on the device, rerun the app, and observe the front-page window.  
The flow is labeled as illegal if nothing unusual is observed, indicating that the app's functionality is not affected. 

\subsection{Learning Methods}\label{sec:classification}
We apply random forest as the supervised learning classifier of FlowIntent since it is commonly used in traffic classification\cite{raghuramu2015uncovering}.
Supervised learning provides promising performance as we show in Section~\ref{sec:super_results}, but it also incurs extra overhead as it requires 
human efforts to filter potential illegal flows generated from expected running instances at the training stage as we discussed above.
Off-the-shell hostname lists of ad or analytics servers may help to identify partial such illegal flows.

In practice, however, IDS may not have the up-to-date hostname list of suspicious servers, and may potentially misclassify illegal location flows generated by ``expected'' location-sharing instances. An alternative approach is to build a model on illegal sharing flows purely from unexpected running instances to generate network-level signature only for privacy leaking flows. We conduct an initial study on the feasibility of unsupervised learning in detecting privacy leakage, by using one-class Support Vector Machine (OCSVM) to derive signatures from the illegal traffic flow found by our user-intention modeling.
OCSVM has been widely applied to anomaly detection recently~\cite{song2013one, jiang2013anomaly, emmott2013systematic}, under the assumption that the majority of the instances in the dataset belong to one class. The evaluation of unsupervised traffic learning is given in Section~\ref{sec:ocsvm}.

\section{Experimental Evaluation}\label{sec:eval}
In this section, we evaluate the performance of FlowIntent, including both the user intention model and the traffic learning models. We note that although we mainly care about the the accuracy of the traffic models in this work, a separate evaluation of user intention modeling is interesting by itself. 

\subsection{Experimental Setup}
We have crawled 10718 Android apps from Google Play~\cite{google-play} and 12480 apps from Baidu App Market~\cite{baidu-app}.
Since we focus on location transmissions through the Internet, we only keep those apps that require both location and network related permissions. 
We have obtained 9055 target apps in total - out of which 6669 apps are from Baidu App Market and 2386 apps are from Google Play.
To collect the dynamic app-level contextual information and the location-sharing HTTP traffic generated by running instances, we run each app inside TaintDroid\cite{enck2014taintdroid} for one minute without intervening the app.
During that period, the UI of the front-page window is recorded using UiAutomator, which is served as the dynamic contextual data of the corresponding instance, 
and all the incoming and outgoing TCP packets are captured with \textit{tcpdump}\cite{tcpdump}. TaintDroid is used to identify the location-sharing traffic flows automatically (using dynamic taint analysis). We have acquired 2803 HTTP location-sharing flows, along with their app-level contexts, from 1626 running instances.

\subsection{User Intention Modeling}\label{sec:eval-ui} 

With the three types of features discussed in Section~\ref{sec:ui}, we apply text classification to build our user intention model for predicting expected and unexpected location sharing behavior at the running instance level. To get the ground truth of the running instances, we manually label each running instance as either ``expected'' or ``unexpected'' regarding sending the location data out of the device based on our understanding of the app functionality after reading its associated app-level context. We have manually selected 634 ``expected'' instances and 634 ``unexpected'' instances from 1626 running instances, and use 10 fold cross-validation on these instances.

Tables~\ref{tab:random_forest} to~\ref{tab:logistic} give the classification results for each of the three classifiers, and Table~\ref{tab:voting} gives the final result after voting is applied. Overall, our model predicts 966 out of the 1002 instances correctly giving a prediction rate of 96.4\%.

\begin{table}
\caption{Text Classification Results and Voting with 10 Fold Cross-validation}
\centering
\subfigure[Random Forest]{
\resizebox{\columnwidth}{!}{
\begin{tabular}{ l | c r }
\hline
& Predicted as illegal & Predicted as legal \\
\hline
Illegal instances & 625 (98.6\%) & 9 (1.4\%)\\
Legal instances & 53 (8.4\%)& 581 (91.6\%) \\
\hline
\end{tabular}
\label{tab:random_forest}
}}\\
\subfigure[Naive Bayes]{
\resizebox{\columnwidth}{!}{
\begin{tabular}{ l | c r }
\hline
& Predicted as illegal & Predicted as legal \\
\hline
Illegal instances & 596 (94\%) & 38 (6\%)\\
Legal instances & 74 (11.7\%)& 560 (88.3\%) \\
\hline
\end{tabular}
\label{tab:bayes}
}}\\
\subfigure[Logistic Regression]{
\resizebox{\columnwidth}{!}{%
\begin{tabular}{ l | c r }
\hline
& Predicted as illegal & Predicted as legal \\
\hline
Illegal instances & 596 (94\%) & 38 (6\%)\\
Legal instances & 70 (11\%)& 564 (89\%) \\
\hline
\end{tabular}
\label{tab:logistic}
}}\\
\subfigure[Voting]{
\resizebox{\columnwidth}{!}{%
\begin{tabular}{ l | c r }
\hline
& Predicted as illegal & Predicted as legal \\
\hline
Illegal instances & 506 (98.7\%) & 7 (1.3\%)\\
Legal instances &  29 (6\%)&  460 (94.0\%) \\
\hline
\end{tabular}
\label{tab:voting}
}}\\
\end{table}

\ignore{
\subsection{Traffic Learning Results}\label{sec:learning_res}
In this section, we describe our learning models for detecting illegal traffic flows, built upon the user intention model and the traffic level features discussed above. We consider two learning models, a supervised learning model that generates signatures for legal location flows, illegal location flows, and non-location flows, and an unsupervised approach that builds a model for illegal flows only. As mentioned earlier, the two approaches provide a different accuracy vs. labeling overhead tradeoff. 
}

\begin{table*}[t]
\centering
\caption{Traffic classification results with user intention learning}
\label{tab:3-class-after-ui}
\small{
\begin{tabular}{ c | c | c | c | c | c }
\hline
Features & TP Rate & FP Rate & Precision  & F-measure  & Attributes with highest information gain\\
\hline
Statistical & 0.847 & 0.076 & 0.848 & 0.844 & \parbox{6cm}{Downlink packet size: mean, max, std. devn,
											Interval between packets: mean,
											TCP packet count} \\
\hline
Lexical & 0.901 & 0.049 & 0.903 & 0.909 & \parbox{6cm}{`map', `loc', `baidu', `jpg', `ads', `weather', `lat', `lng'}\\
\hline
Both & 0.911 & 0.045 & 0.913 & 0.911 \\
\hline
\end{tabular}
}
\end{table*}

\begin{table}[h]
\caption{Traffic classification results with true instance label}
\label{tab:3-class-none-ui}
\small{
\resizebox{\columnwidth}{!}{%
\begin{tabular}{ c | c | c | c | c }
\hline
Features & TP Rate & FP Rate & Precision  & F-measure\\
\hline
Statistical & 0.856 & 0.072 & 0.857 & 0.856   \\
Lexical & 0.921 & 0.039 & 0.923 & 0.921 \\
Both & 0.926 & 0.037 & 0.928 & 0.926 \\
\hline
\end{tabular}
}
}
\end{table}

\begin{table}[h]
\caption{Prediction Results: TP and FP rate is calculated for one class against all other classes}
\label{tab:4-predic-result}
\small{
\resizebox{\columnwidth}{!}{%
\begin{tabular}{ c | c | c | c | c }
\hline
App Class & TP Rate & FP Rate & Precision  & F-measure\\
\hline
Non-loc & 0.96 & 0.067 & 0.88 & 0.92   \\
Illegal-loc & 0.89 & 0.052 & 0.89 & 0.89 \\
Iegal-loc & 0.88 & 0.018 & 0.96 & 0.92 \\
\hline
\end{tabular}
}
}
\end{table}

\begin{table}[t]
\caption{Prediction results on non-location flows reported by TaintDroid}
\label{tab:5-taint-miss}
\resizebox{\columnwidth}{!}{%
\begin{tabular}{ c | c | c | c | c }
\hline
& Legal loc flows & Illegal loc flows & Non-loc flows & Unknown \\
\hline
Predicted as legal & 311  & 23 & 16 & 38  \\
Predicted as illegal & 17 & 185 & 11 & 21\\
\hline
\end{tabular}
}
\end{table}

\subsection{Supervised Traffic Learning}\label{sec:super_results}
Given the labeled running instances, we then identify (a part of) illegal flows from the ``expected'' instances using the lists of ad and analytics servers given in~\cite{crussell2014madfraud}. We have thus obtained 896 flows from 467 ``expected'' running instances, and 817 flows from 535 ``unexpected'' running instances, with 319 different server hostnames in total. As we discussed in Section~\ref{sec:dataset-flow}, some of the flows in the former set may actually be illegal, and we allow this noise in the training dataset. For the testing purpose, however, we have manually identified 43 illegal flows from the 896 location sharing flows generated by the ``expected'' instances, using the approach discussed in Section~\ref{sec:dataset-flow}. We have also collected 850 non-location flows from apps that do not ask for location permission. We then apply the random forest method on these traffic flows to construct our supervised learning model.

Given TP = number of true positives, FN = number of false negatives, FP = number of false positives and TN = number of true negatives, the efficiency of prediction of the model is measured based on the following characteristics:\newline
TP Rate $= \frac{TP}{TP+FN}$, \quad FP Rate $= \frac{FP}{FP+TN}$, \newline
Precision $= \frac{TP}{TP+FP}$, \ \ F-measure $= \frac{2TP}{2TP+FP+FN}$.





The results of our model with 10-fold cross-validation are given in Tables~\ref{tab:3-class-after-ui}-\ref{tab:4-predic-result}. Table~\ref{tab:3-class-after-ui} shows the prediction result when the instances are labeled using our user intention model, while Table~\ref{tab:3-class-none-ui} gives the result when true instance labels are used in the training of the traffic model. Table~\ref{tab:4-predic-result} gives the prediction result for each type of flows that correspond to Table~\ref{tab:3-class-after-ui}. We make the following observations.
\begin{itemize}
\item First, the traffic classification model achieves 91.1\% F-measure by using both statistical and lexical features, even when some running instances are potentially mislabeled due to the inaccuracy of the user intention model.  When the true running instance classes are used, the F-measure increases to 92.6\%.
The fact that the amount of illegal flows from expected instances is relatively small compared to the legal flows also contributes to the high precision.
Therefore, our user intention modeling only incurs a slight loss in accuracy, while saving the effort of manually labeling a large number of instances.
\item Second, lexical features alone can provide relatively good predication accuracy, which can be further improved by including statistical features.
Among the set of most useful lexical features shown in Table~\ref{tab:3-class-after-ui}, `loc', `jpg', `lat', `lng' are useful in distinguishing location and non-location flows, while the rest can be used to distinguish all the three types of traffic. In particular, `baidu' is a good indicator of legal flows because most benign instances in China use the map service provided by Baidu.
\item Third, we observe that the F-measure results of using statistical features only are also good. This indicates that our approach can be potentially extended to HTTPS flows as well, even though the lexical features cannot be applied to HTTPS traffic.
The set of statistical features with the highest information gain as shown in Table~\ref{tab:3-class-after-ui} is consistent with our observation in Section~\ref{sec:features}.
\end{itemize}

\vspace{0.5ex}
\noindent{\bf Improved Location Flow Detection:}
In addition to achieving a promising detection accuracy, our learning model is able to detect new location sharing flows that are undetectable by TaintDroid, which highlights the advantage of having network level signatures. To confirm this, we randomly select 5510 flows generated by 670 running instances that TaintDroid does not report location sharing. Among these flows, our model detected 622 of them to be location related, and 388 of them are classified as legitimate, and the rest are classified as illegal. We then check the ground truth for each of these flows manually.
The result is shown in the Table~\ref{tab:5-taint-miss}. In the table, the `Unknown' column indicates the cases where we cannot identify the ground truth, when the traffic is encrypted and the URLs are not familiar. Our model is able to detect 311 legal location flows and 185 illegal flows correctly, all of which are missed by TaintDroid. The result shows that our learning model is able to identify location sharing flows that are missed by host-based taint analysis, which strongly indicates the benefit of considering network-level features.


\subsection{Unsupervised Traffic Learning}\label{sec:ocsvm}
To reduce human intervention in identifying true legal flows, unsupervised learning model is built on illegal sharing flows purely from unexpected running instances to generate network-level signature.
%
From 817 flows generated by 535 ``unexpected'' instances, we randomly sample 654 of them as the training data.
We apply the Support Vector Data Description~\cite{svdd} to identify a minimum boundary of the dataset.


We use 160 true illegitimate flows that are unused during the training stage as one testing dataset.
We also randomly sample 160 \textit{other} flows from the pool of legal location transmissions and non-location transmissions to build another testing dataset.
The results are shown in Table~\ref{tab:ocsvm-results}.
The model achieves high true positive rate (94.4\%) at the cost of the precision (83.4\%).
We finally acquire F-measure of 88.6\%.
As we expect, the overall performance of unsupervised learning model is a bit lower compared to the supervised model shown in the last subsection.
However, the advantage of less human intervention makes the unsupervised model to be easier to deploy in some practical scenarios.

\begin{table}[t]
\caption{Prediction results of OC-SVM Model}
\label{tab:ocsvm-results}
\centering
\small{
\begin{tabular}{ c|  c | c }
\hline
& Illegal loc flows & Others \\
\hline
Predicted as illegal & 151 & 30  \\
Predicted as others & 9 & 130 \\
\hline
\end{tabular}
}
\end{table}

\section{Conclusion}\label{sec:conclusion}
In this paper, we develop FlowIntent, a proof-of-concept system that makes the first attempt to identify the location-leaking traffic flows from the mismatch between user intention and network behavior.
Compared to system-level detection approaches, our network-level signatures are easier to deploy at Intrusion Detection Systems to monitor a large number of devices simultaneously, while introducing zero overhead at the end hosts. FlowIntent also captures the sensitive transmissions missed by the state-of-the-art dynamic taint analysis systems. In contrast to previous network-level detection techniques that rely on a given set of malicious domain names, FlowIntent can better adapt to the fast growth of app market and new leakage patterns through user-intention modeling. 
We have built our learning models using 1002 location sharing instances identified from more than 20,000 apps crawled from Android app markets. Our approach achieves about 91\% accuracy in distinguishing between legitimate and illegitimate location transmissions.

\section*{Acknowledgements}
The effort described in this article was partially sponsored by the U.S. Army Research Laboratory Cyber Security Collaborative Research Alliance under Contract Number W911NF-13-2-0045.  The views and conclusions contained in this document are those of the authors, and should not be interpreted as representing the official policies, either expressed or implied, of the Army Research Laboratory or the U.S. Government. The U.S. Government is authorized to reproduce and distribute reprints for Government purposes, notwithstanding any copyright notation hereon.

\bibliographystyle{abbrv}
\bibliography{ref}

\end{document}